\documentstyle[11pt]{article}

\begin{document}

\title{\center{ULTRAVIOLET SKY SURVEYS \\ {\small
 Invited review for IAU 179 Symposium on Multi-wavelength Sky Surveys}}}


\author{N. Brosch \\
Dept. of Astronomy and Astrophysics and the Wise Observatory, \\
Raymond and Beverly Sackler Faculty of Exact Sciences, \\
 School of Physics and Astronomy, Tel Aviv University,
Tel Aviv 69978, Israel }

\maketitle


\section{Introduction}

   Among all spectral bands, the ultraviolet has long been neglected,
      despite the  advantage of small space experiments: the sky is very dark, 
thus detection of faint
    objects does not compete against an enhanced background (O'Connell
    1987) and the telescope construction techniques are very similar
    (at least longward of $\sim$50 nm) to those of optical astronomy. 

     The short history of UV astronomy can be
    divided into two eras, until the flight of TD-1 and since the availability of the
    TD-1 all-sky survey. Very little was
    accomplished in terms of general sky surveys during this second era. 
         The UV domain may be divided into the "regular" ultraviolet,
     from shortward of the spectral region
    observable from ground-based observatories ($\sim$320 nm) to  
    below the Lyman break at $\sim$80 nm, and the region from the Lyman
    break to the fuzzy beginning of the X-ray domain,
 arbitrarily defined as $\sim$6 nm$\approx$200 eV). 
The first segment is called "UV" and the second "extreme UV" (EUV).
    Observational techniques used in the
    EUV are more similar to those in X-ray astronomy, whereas the UV is
      more like the optical. 
 
The units used here are "monochromatic magnitudes",
    defined as: 
\begin{equation}
    m(\lambda)=-2.5 \, log[f(\lambda)]-21.175 
\end{equation}
    where f($\lambda$) is the source flux density in erg/sec/cm$^2$/\AA\, 
    at wavelength $\lambda$.
    The background brightness is described in "photon
    units" (c.u.=count units) which count the photon flux in a
    spectral band, per cm$^2$, per steradian, and per \AA\,. At 150 nm, 
    1 c.u.=1.32 10$^{-11}$ erg/cm$^2$/sec/\AA\,/steradian, or 1.32 10$^{-13}$
    W/m$^2$/nm/steradian, or 32.6 mag/square arcsec. 

    Although only few missions performed full sky surveys in the UV or EUV,
    many scanned or imaged restricted sky regions
     and provided  information about the
    deeper UV sky.  O'Connell (1991) reviewed UV imaging experiments and their results
    updated to 1990.

    In parallel with the development of UV astronomy,  first steps
    were taken to study the EUV sky. Detections in the EUV range are hampered by the 
    opacity of the interstellar medium
    (ISM). From 91.2 nm shortward to about 10 nm the opacity is high,
    because of the photoelectric cross-section of H$^0$, and to a lesser
    extent of He$^0$ (below 50.4 nm) and He$^{+1}$ (below 22.8 nm). 

    The first studies in the EUV range were with
    rocket-flown instruments (Henry  {\it et al.}  1975 a, b, c), which measured a
    few very bright sources and established calibrators. The earliest
    observations below Lyman $\alpha$ were by Belyaev  {\it et al.}  (1971).  
    The culmination  was the EUV instrument
    flown on the Apollo-Soyuz mission in 1975, when four EUV point sources were
    discovered (Lampton  {\it et al.}  1976, Margon  {\it et al.}  
    1976, Haisch  {\it et al.}  1977, Margon  {\it et al.}  1978).   
    The Voyager spacecraft explored the EUV sky with their Ultraviolet
    Spectrometers (UVS: Sandel, Shemansky \& Broadfoot 1979). For a number
    years the two Voyager spacecraft were the most distant
    astronomical observatories (Holberg 1990, 1991).

     \section{ The TD-1 Era }
 
    Modern UV astronomy began with the first UV all-sky
    survey 
    by the ESRO {\bf TD-1} satellite, described by Boksenberg  {\it et al.}  (1973). 
 The   all-sky catalog of UV sources was published by Thompson
     {\it et al.}  (1978) with 31,215 stars with S/N$>$10 in
    all four TD-1 bands. An unpublished version, with lower S/N,
     has 58,012 objects. 
    The TD-1 S2/68 experiment is a benchmark
    against which all other sky surveys are and will be measured. 
   
    After TD-1, the various UV and EUV efforts can be characterized as
    either imagers or spectrometers. Among the imagers, some were orbiters
    and others were on short-duration flights. Some major missions were ANS and IUE.
     {\bf ANS}  was described by Van Duinen  {\it et al.}  (1975),
    Wesselius  {\it et al.}  (1982), and de Boer (1982). One of the greatest 
    successes of any orbiting astronomical
    instrument was the {\bf IUE} observatory (Boggess {\it et al.}
    1978).  The IUE data  are a valuable resource, mainly
    after the final reprocessing of all the low-dispersion spectra into the
    final Uniform Low-Dispersion Archive (ULDA).  
    
    The NRL experiment {\bf S201} was described by Page  {\it et al.}  (1982)
and operated automatically on the Moon
    during the Apollo 16 mission in April 1972.  
    Ten  20$^{\circ}$   diameter fields were observed,
    the experiment covering $\sim$7\% of the sky. The results 
   were discussed by Carruthers \& Page (1976, 1983, 1984a, 1984b).  
    Two experiments flew on sounding rockets.
    {\bf GUV} flew on 21 February 1987 and is described by Onaka  {\it et al.}  (1989). 
    The GUV observations were re-analyzed by Kodaira  {\it et al.}  (1990).
 The {\bf UIT prototype} flew on a number of rocket flights and with different 
  focal plane assemblies. Bohlin  {\it et al.}  (1982) described the instrument and its 
    observations of the Orion nebula. Other observations  are described by 
   Smith \& Cornett (1982),  Smith  {\it et al.} (1987), and Bohlin {\it et al.}  (1990).
    
   UV observations from balloon-borne telescopes at 40+ km were performed by a   
  collaboration between the Observatoire de
    Geneve and the Laboratoire d'Astrophysique Spatiale  of
    Marseille. The  stabilized gondola (Huguenin \&
    Magnan 1978) carried telescopes tuned for imaging observations in a bandpass 
centered at $\sim$200 nm and   $\sim$15 nm wide.
  {\bf SCAP-2000} was described by Laget (1980), Donas  {\it et al.}  (1981),
    and Milliard  {\it et al.}  (1983). The   results 
 were published by Donas
     {\it et al.}  (1987) and Buat  {\it et al.}  (1987, 1989).
  {\bf FOCA} is a 39 cm diameter telescope  (Milliard  {\it et al.} 1991)
 which surveyed some 70  degrees$^2$ of the sky to $\sim$19 mag.
    Results   were reported by
    Laget  {\it et al.}  (1991a, 1991b), Vuillemin  {\it et al.}  (1991), 
    Courtes  {\it et al.} (1993), Buat  {\it et al.} (1994), Bersier {\it et al.}
    (1994), Reichen  {\it et al.}  (1994), Donas {\it et
    al.} (1995), and Petit  {\it et al.}  (1996).
    Galaxy counts and color distributions for objects in the  range
    15.0-18.5 mag were published by Milliard  {\it et al.}  (1992) and used to
 predict  UV galaxy counts (Armand \& Milliard  1994). 

    The {\bf Wide-Field UV Camera} flew in December 1983 on the Space Shuttle
    and produced some very wide-field UV images ( Courtes  {\it et al.} 1984).
    The NRL group headed by Carruthers flew a number of far-UV wide-field imagers
    on rockets (Carruthers  {\it et al.}   1980).
 These flights used the Mark II {\bf FUVCAM}  
    (Carruthers {\it et al.} 1993, 1994) which flew on the Space Shuttle
    in spring 1991. The results were published in a series of
    papers dealing with individual fields (Schmidt \& Carruthers 
    1993a, 1993b, 1995).  
    The 40 cm telescope {\bf GLAZAR} operated on the Mir space station
    (Tovmassian  {\it et al.} 1988,  1991a, 1991b) and 
    results   were reported by Tovmassian  
   {\it et al.}  (1993a, 1993b, 1994, 1996a).  

    {\bf FAUST} is the Fusee Astronomique pour l'Ultraviolet Spatiale, or the Far
    Ultraviolet Space Telescope, first described by
    Deharveng  {\it et al.}  (1979). On SPACELAB-1 in 1983
FAUST did not obtain significant data
because of high on-orbit background (Bixler  {\it et al.}  1984). 
    During the second flight, on board
    the Shuttle Atlantis in March 1992 (Bowyer {\it et al.}
    1993), FAUST observed 22 regions $\sim8^{\circ}$ in diameter and produced a 
    catalog of 4,698 UV sources (Bowyer {\it et al.} 1994a). Selected 
results from the FAUST imagery are by
    Deharveng  {\it et al.} (1994), Haikala  {\it et al.}  (1995), and  
    Courtes  {\it et al.}  (1995).   
    A program of systematic investigation of FAUST images takes place the
    Tel Aviv University and include optical observations from the Wise
    Observatory. To date, we analyzed completely four FAUST fields, the
    North Galactic Pole (Brosch  {\it et al.}  1996a) and three fields covering 
      Virgo (Brosch  {\it et al.}  1996b). 
 
    The Ultraviolet Imaging Telescope ({\bf UIT}) was described by Stecher {\it et al.}
    (1992). It flew on the
    Space Shuttle during the ASTRO-1 (December 1990) and ASTRO-2 flights
    (March 1995).  First results were published 
    in a dedicated publication
   (1992 Astrophys. J. Lett. {\bf 395}). Other results are in   
    Hill  {\it et al.}  (1993, 1994, 1995a, 1995b, 1996).
    The UIT source  catalog
    (Smith  {\it et al.}  1996) covers 16  degrees$^2$ of the sky and contains 2244
    objects from 48 pointings in the ASTRO-1 flight.
 
    Attempts to measure the diffuse UV background   
    consist of many observations with wide-field instruments. Notable among these
    are the  two Shuttle-borne
    UVX instruments from JHU and Berkeley (Murthy  {\it et al.}  1989; Hurwitz {\it et al.}
    1989). In addition, observations done for other purposes were used to  
    derive
    the UV background, {\it e.g.,} Waller  {\it et al.} (1995).
    
    \section{ Modern EUV observations }

    The EUV sky was explored by the
    {\bf EUV Wide Field Camera} on the ROSAT X-ray
    all-sky survey satellite described by Pounds \& Wells (1991). The first EUV
    all-sky survey was during 1990-1991 (Pye
    1995).  Initial results were
    reported by Pounds  {\it et al.}  (1993) as the WFC Bright Source Catalog (BSC).
    The reprocessed data make up the 2RE catalog (Pye {\it et al.} 1995)
    with 479 EUV sources. In 2RE
     52\% of the sources are active F, G, K,
    and M stars, 29\% are hot white dwarfs, and less than 2\% are AGNs.   

    The region bordering the EUV and the X-rays was explored by the {\bf ALEXIS}
    spacecraft (Priedhorsky 1991).  The first sky maps  were
    produced on 4 November 1994. The ALEXIS team calculated that $\sim$10\%
    of the brightest EUVE sources (see below) should be detectable.
    Most sources are WDs; the
    catalog from the first three years of operation will probably contain $\leq$50
    sources.

    The EUV sky was   investigated by the Extreme Ultraviolet
    Explorer ({\bf EUVE}) spacecraft (Bowyer \& Malina 1991). EUVE mapped the sky
    in four spectral bands, from 7 to 70 nm (18 to 170 eV). The first
    results were published as "The First EUVE Source Catalog" (Bowyer {\it et al.}
    1994) with 410 sources. The Second EUVE Source Catalog (2EUVE) has
    recently been published (Bowyer  {\it et al.}  1996).  
    The majority of the identified sources in 2EUVE (55\%) are G, K, and M stars.  
 
    A new catalog,  to $\sim$60\% of the thresholds of the second
    EUVE catalog, has been produced by Lampton {\it et al.} (1996) with
    534 coincident sources between the EUVE 10 nm list and the ROSAT
    all-sky survey sources detected in the broadband event window
    (0.1-2 keV), of which 166 were not
    previously discovered. Of these, 105 have been identified and 77\%
    of them are late-type stars. White dwarfs and early-type stars
    make up only ~14\% of the sources, and there are no extragalactic
    objects at all.
 
        \section{Comparison of Survey Missions }

    The various missions surveying the UV sky can be compared in terms of a
    "power" parameter $\theta$, introduced by Terebizh (1986), used by Lipovetsky
    (1992) in a comparison of optical surveys, and slightly modified here:
\begin{equation}
    \theta=\frac{\Omega}{4 \pi}  10^{0.6*(m_L-10)} 
\end{equation}
    where $\Omega$ is the sky area covered by the survey (4$\pi$ for TD-1)
    and m$_L$ is the limiting magnitude of the survey. 

The various parameters relevant to the missions discussed here are collected
in Tabele 1. An all-sky
    survey to m$_L\approx$8.5 (such as TD-1) 
    has the same
    "survey power" as one HST WFPC-2 image exposed to show m$_{UV}$=21 objects. 
    Because of this, and because not all surveys cover the entire sky, it may be
    more useful to look at another estimator, the density of sources
    detected (or which are expected to be detected) by a certain experiment. 
This estimator indicates that the field of UV astronomy retains its vitality; 
the source density increases exponentially with time.

\begin{table}[htb]
\begin{center}
\caption{ UV and EUV survey missions}
    \begin{tabular}{cccrrrrl} 
\hline
    Mission & Year & $\Omega$ (ster)& m$_L$ & $\theta$ & $\lambda\lambda$ (nm) & N$_{sources}$ & Notes   \\ \hline 
    TD-1    & 1968-73 & 4$\pi$ &  8.8 & 0.19 & 150-280 & 31,215 &  1 \\ 
    S201     & 1972 & 0.96  & 11  & 0.30 & 125-160 & 6,266 & \\ 
    WF-UVCAM & 1983 & 1.02  & 9.3 & 0.03 & 193     & ?    & \\ 
    SCAP-2000 & 1985 & 1.88 & 13.5 & 18.9 & 200 & 241 & 2 \\ 
    GUV       & 1987 & 5 10$^{-3}$ & 14.5 & 0.2  & 156 & 52  &  Pointed phase  \\ 
   GSFC CAM & 1987+ & 0.03 & 16.3 & 14.4 & 242 & $\sim$200 & Virgo observation \\ 
    FOCA & 1990+ & 0.02 & 19 & 377 & 200 & $\sim$4,000 & Estimated \\ 
    UIT-1 & 1990 & 3.8 10$^{-4}$ & 17 & 0.48 & $\sim$270 & 2,244 & UIT catalog \\ 
    GLAZAR & 1990 & 4.4 10$^{-3}$ & 8.7 & 6 10$^{-4}$ & 164 & 489 &  \\
    FUVCAM & 1991 & 0.09 & 10 & 7.5 10$^{-3}$ & 133, 178 & 1,252 &  3 \\ 
    FAUST & 1992 & 0.33 & 13.5 & 3.3 & 165 & 4,698 & \\ 
    UIT 1+2 & 1990, 95 & 1.3 10$^{-3}$ & 19 & 26 & 152-270 & 6,000 ? & 4\\  
\hline
    HST WFPC & 1990+ & 3.9 10$^{-4}$ & 21 & 123 & 120-300 & 50,000 ? & 5 \\ \hline 
    MSX UVISI & 1997+ & 4$\pi$ ? & 13.9 & 218 & 180-300 & ? & \\ 
    GIMI      & 1997+ & 4$\pi$   & 13.6 & 136 & 155 & 2.5 10$^5$ & 6 \\ 
    TAUVEX & 1998+ & 0.06 & 19 & 11,700 & 135-270 & 10$^6$ & 7 \\ 
     \hline
WFC & 1992 ? & 4$\pi$ & - & - & 10, 16 & 479 & \\
ALEXIS & 1994+ & 4$\pi$ & - & - & 13-19 & 50? & \\
EUVE & 1992 ? & 4$\pi$ & - & - & 7-70 & 734 & 8 \\
\hline
    \end{tabular}
\end{center}

    Notes to Table 1: \\
      1: The unpublished extended version has 58,012 sources. \\
	2: 92 stars (Laget 1980) and 149 galaxies (Donas {\it et al.} 1987). \\
     3: Only the Sag and Sco fields (Shuttle flights) included. \\
 4: Assumes 66 pointings for ASTRO 1 and 100  for ASTRO 2.\\
    5: Assumes 1000 observations with HST with UV filters on WFPC-2. \\
    6: Assumes  2$\times$ stars per magnitude w.r.t. TD-1. \\
    7: Assumes 5000 independent pointings to end-of-life. \\
    8: Number of sources in the 2nd EUVE catalog. 
 \end{table}

    \section{The Resultant Sky Picture }

     The combined results yield a picture in which most of the stars
    detected by TD-1, FAUST, SCAP and FOCA are early-type B, A
    and F. However, most of the stars included in the UIT catalog are probably
    late-type (G and later). In the FAUST fields   where
    the reduction and identification processes are complete, we find
    almost equal fractions of A-F stars (70  and 75\%).
    Except for the fields studied at Tel Aviv (Brosch {\it et 
    al.} 1995, 1996), most surveys used exclusively correlations with existing
    catalogs to identify sources. These sometimes mis-identify objects, as
     some likely early-type stars are just below the catalog thresholds.
     
    The UV information on galaxies is very sparse and   a  statistically complete
    sample of a few 1000's galaxies is lacking. In the absence of very deep 
surveys in more than a single spectral
    band,  our
    information about a significant number of galaxies originates from  
    SCAP-2000 (Donas  {\it et al.}  1987) and FOCA (Milliard  {\it et al.}  1992).
    These measurements consist of integrated photometry at 200 nm of a few
    hundred galaxies. In  the range
    16.5-18.5 mag galaxies  dominate the source counts at high $\mid$b$\mid$. These 
have B=18-20 and 
   [2000-V]$\approx-$1.5. 
    Using the "field" galaxy luminosity function in the UV from 
    Deharveng  {\it et al.}  (1994) the differential number density is:
\begin{equation}
log N(m)=0.625 \times m_{200}-9.5
\end{equation}

    Studies by UIT and FAUST emphasize the importance of the dust in
    understanding the UV emission.  
    Bilenko \& Brosch (1996) analyzed the TD-1 catalog and a version of the
    Hipparcos Input Catalog transformed to the TD-1 bands
      and showed that
    the UV exinction is very patchy, with different  
    extinction gradients on scales  $<10^{\circ}$.
    Tovmassian  {\it et al.}  (1996b) used  GLAZAR observations of a 12
    degree$^2$ area in Crux to establish that the dust distribution  
    is very patchy, with most of the space relatively clear of dust. 
     The EUVE catalogs
    (Bowyer  {\it et al.}  1994, 1996; Lampton  {\it et al.}  1996) 
    confirm the previously known
    features of the local ISM (a "tunnel" to CMa with very low HI column
    density to ~200 pc. and close to the Galactic plane, a cavity connected
    with the Gum Nebula in Vela, a shorter 100 pc. tunnel to 36 Lynx, and the
    very clear region in the direction of the Lockman hole).

    The accurate measurement of the UV sky background (UVB), with the expectation
    that it could set meaningful cosmological limits, has been the goal of
    many rocket, orbital, and deep space experiments. Observational
    results were summarized by Henry (1982), Bowyer
    (1990), Bowyer (1991), and Henry (1991). 
    The various origins of the   UVB can be separated
    into "galactic" and "high latitude". The latter is an $\sim$uniform
    pedestal, onto which the former is added in various amounts
    depending on the direction of observation. The "galactic" component
    can be $\sim$one order of magnitude more than the "high latitude"
    one. Most is probably light scattered off
    dust particles in the ISM and the rest is from the gaseous component
    of the ISM (HII two photon emission and H$_2$ fluorescense in
    molecular clouds). The  "high latitude" component is
    also mostly galactic,  light scattering off
    dust clouds at high $\mid$b$\mid$.

    The extragalactic component of the UVB (eUVB)
   can be at most 100-400 c.u. (Murthy \& Henry 1995). 
    Whenever N(HI)$>$2 10$^{20}$ cm$^{-2}$, the main
    contributor is dust-scattered starlight. The
    low level  eUVB is  probably 
    integrated light of galaxies (Armand  {\it et al.}  1994), or
     Milky Way light scattered off dust grains in
    the Galactic halo (Hurwitz  {\it et al.}  1991), or intergalactic Ly$\alpha$
    clouds  contributing their
    recombination radiation (Henry 1991).
 
    The low  UVB away from orbital and galactic 
    contaminants has recently been confirmed by  
     ASTRO-1 UIT images (Waller  {\it et al.}  1995). After correcting
    for orbital background and zodiacal light, and after accounting for 
    scattered Galactic light by ISM cirrus clouds (from the IRAS 100
    $\mu$m emission), the extrapolated  UV-to-FIR correlation
    to negligible FIR emission indicates  [eUVB]$\approx$200$\pm$100 c.u.

    The shorter wavelength UVB has been observed with the Voyager UVS  
down to 50 nm (Holberg 1986).
    A very deep EUVE spectroscopic observation of a large region on
    the ecliptic has recently been reported (Jelinsky  {\it et al.}  1995)
but  the only emission lines observed were He I and He II  
    (58.4, 53.7, and 30.4 nm), which originate from scatted Sunlight by
    the geocoronal and/or interplanetary medium and no continuum was detected.  

    The "true" eUVB can be evaluated from the FOCA
data (Milliard  {\it et al.}  1992).
    The  galaxy counts, for  15.0$\geq m_{200}\geq$18.5, 
      extrapolated to m$_{200}$=20.0, give a contribution of 
    $\sim$100 c.u.'s from only UV galaxies.
 Milliard (1996, private communication) studied the nature of sources
in the A2111 FOCA field. The majority are 
emission-line galaxies, only 16\% in A2111 and the rest 
are in the foreground or background up to z$\approx$0.7.  

 The detection and identification of faint UV galaxies may help
resolve the issue of a real eUVB. Recent indications
are that the eUVB  is negligible. Waller {\it et al.}
(1995) estimated 200$\pm$100 c.u. in the UIT near-UV band. 
Unpublished results from an analysis of 17 years of Voyager UVS
spectra (Murthy {\it et al.} 1996, in preparation) indicate that at $\sim$100 nm the UVB
is $\leq$100 c.u. (1$\sigma$). However, when adopting the 
FOCA galaxy counts and extrapolating  to m$_{UV}$=20 and to $\sim$1000\AA\, using
the SEDs of starburst galaxies from Kinney {\it et al.} 
(1996), the 
contribution due to sources unresolved by UVS violates the Murthy {\it et al.}
limit. An extrapolation to m$_{UV}$=23 violates also the UIT constraint. It is
therefore necessary to investigate the faint end of the UV galaxy distribution
to understand the nature and reality of the eUVB.

    \section{The future }

    Two UV missions are approved, funded, built, and integrated into their
    carrier spacecraft. These are the UVISI on MSX, and GIMI on ARGUS, which
 will produce full or partial UV
    sky surveys. MSX was launched on 24 April 1996 and the
    operation of UVISI is expected to start in 1997. ARGUS has been slightly
    delayed and will be launched in 1997. 
 
The narrow field UV imager of {\bf UVISI} (Heffernan
     {\it et al.}  1996)  is
    sensitive to sources which produce 2 photons/cm$^2$/sec, 
    {\it i.e.,} m$_L\approx$13.9 (monochromatic, at 240 nm). It is not clear 
how much of the sky will 
    MSX survey. However, {\bf GIMI} on ARGOS has as a declared goal 
    the production of a full sky survey in three UV bands. The most recent 
    description of GIMI (Carruthers \& Seeley 1996) indicates m$_L\approx$13.6.

    TAUVEX, {\bf T}el {\bf A}viv University {\bf UV Ex}plorer, (Brosch  {\it et al.}  1994) 
    is the most
    advanced attempt to design, build and operate a flexible instrument for
    observations in the entire UV band. TAUVEX images the same 0$^{\circ}$.9 
    FOV with three co-aligned telescopes with an image quality of about
    10". It is part of the scientific complement of the
    Spectrum X-$\gamma$ spacecraft. The projected performance is detection
    of objects 19 mag and brighter with S/N$>$10  in three
    bands $\sim$40 nm wide, after a four hour pointing. At high
   $\mid$b$\mid$, each pointing is expected to result in the
    detection of some tens of QSOs and AGNs (mainly low-z objects) and some
    hundreds of galaxies and stars. A three-year operation will cover $\sim$5\%  
   of the sky to  m$_{UV}\approx$19 mag. 
 
    The deepest observations of the UV sky   
     should be made far from the Earth's geocorona, away from the Sun, and
   out of the ecliptic.   Multi-purpose missions to the outer planets 
 could be used for astronomy during their cruise phase.
    Much cheaper options are UV observations from  long-duration super-pressure 
balloons at 40-45 km altitude, for flights of
    weeks to months, with real-time operation 
      using the TDRSS data relay satellites (Israel 1993).  
    The reactivation of the GLAZAR telescope should be considered. 
An interesting possibility is to conduct a deep UV survey of a 
    very small fraction of the sky with the HST, to learn about the faintest
    UV sources, beyond the capability of UIT, FOCA, or TAUVEX. 
 
   \section{Conclusions}

A  review of the  UV and EUV sky reveals that while most UV sources are early-type 
stars, most EUV sources are late-type stars. It is clear that UV and EUV observations 
are hampered by ISM opacity.  Very faint UV sources are mostly galaxies, and the UVB, 
if it exists, is very faint and may be fully explained 
by galaxies. 

    Significant improvement in our knowledge of the
     UV sky can be achieved through judicious use of
     multi-purpose platforms, where the UV science piggy-backs
    on other spectral ranges.  The heritage of
    past missions should be fully realized before embarking on new
    adventures. In this light, the extension of the EUVE mission in a
    low-cost mode is commendable; the extension of the right-angle surveys
    should bring in  more faint EUV sources.

    What science needs is either a multi-spectral all-sky UV imaging survey  to 19-20
    monochromatic magnitude, or a combination of cheaper alternatives, including a  
    long-duration, very high-altitude balloon with a FOCA-type
    telescope and with a high-resolution detector with electronic readout 
    for a sky survey in the 200 nm band, combined with the reactivation of 
    the GLAZAR telescope after the status of its
    optics has been reassessed, with a modern, digital, photon-counting
    detector, and a mini-survey with HST in the UV. Additionally, an extended 
     EUV all-sky survey,  $\sim100\times$ more
     sensitive than EUVE and with better angular resolution, is still required.
 
    \section*{Acknowledgements}

    UV research at Tel Aviv University is supported by special grants from
    the Israel Space Agency, from the Austrian Friends of Tel Aviv
    University, and from a Center of Excellence Award from the Israel
    Academy of Sciences. I acknowledge support from a US-Israel Binational
    Award to study FAUST UV sources, and the
    hospitality of NORDITA and the Danish Space Research Institute where
    parts of this review were prepared. I am grateful for the help of
   David Bersier, Benny Bilenko,  Jeff Bloch,  Jeff Hill, David Israel,
Michael Lampton,  and Bruno  Milliard in preparing this review.


\newpage

 \begin{thebibliography}{}

\bibitem{}    Armand, C. \& Milliard, B. 1994 {\it Astron. Astrophys.} {\bf 282}, 1.  

\bibitem{}    Belyaev, V.P. {\it et al.} 1971 {\it Cosmic Res.} {\bf 8}, 677. 

\bibitem{}    Bersier, D. {\it et al.} 1994 {\it Astron. Astrophys.} {\bf 286}, 37. 

\bibitem{}    Bilenko, B. \& Brosch, N. 1996 preprint. 

\bibitem{}    Bixler, J. {\it et al.} 1984 {\it Science} {\bf 225}, 184.

 \bibitem{} Boggess, A.  {\it et al.}   1978 {\it Nature} {\bf 275}, 372.

\bibitem{}    Bohlin, R.C. {\it et al.} 1982 {\it Astrophys. J.}
    {\bf 255}, 87.

\bibitem{}    Bohlin, R.C. {\it et al.} 1990 {\it Astrophys. J.} {\bf 352}, 55.

\bibitem{}    Boksenberg, A. {\it et al.} 1973 {\it Mon. Not. R. astr. Soc.} {\bf 163}, 291.

\bibitem{}    Bowyer, S. 1990 in "The Galactic and Extragalactic Background Radiation"
    (S. Bowyer \& Ch. Leinert, eds.), Dordrecht: Reidel, p. 171.

\bibitem{}    Bowyer, S. 1991 {\it Ann. Rev.  Astron. Astrophys.} {\bf 29}, 59.

\bibitem{}    Bowyer, S. \& Malina, R.F. 1991 "Extreme Ultraviolet Astronomy"
		  (R.F.Malina \& S. Bowyer, eds.), New York: Pergamon
		  Press, p. 397.

\bibitem{}    Bowyer, S. {\it et al.} 1993 {\it Astrophys. J.} 
{\bf 415}, 875 
 
\bibitem{}    Bowyer, S. {\it et al.} 1994a {\it Astrophys. J. 
Suppl.} {\bf 96}, 461.

\bibitem{}    Bowyer, S. {\it et al.} 1994 {\it Astrophys. J. Suppl.} {\bf 93}, 569. 

\bibitem{}    Bowyer, S. {\it et al.}
    1996 {\it {\it Astrophys. J.} Suppl.} {\bf 102}, 129.

\bibitem{}    Brosch, N. 1991  {\it Mon. Not. R. astr. Soc.} {\bf 250}, 780.

\bibitem{}    Brosch, N. {\it et al.}
    1994   {\it SPIE} {\bf  2279},   p. 469.

\bibitem{}    Brosch, N. {\it et al.} 1995 {\it Astrophys. J.} {\bf 450}, 137. 

\bibitem{}    Brosch, N.  {\it et al.}  1996 preprint. 

\bibitem{}    Buat, V. {\it et al.} 1987 {\it Astron. Astrophys.} {\bf 185}, 33. 

\bibitem{}    Buat, V. {\it et al.} 1989 {\it Astron. Astrophys.} {\bf 223}, 42. 

\bibitem{}    Buat, V. {\it et al.} 1994
    {\it Astron. Astrophys.} {\bf 281}, 666.
 

\bibitem{}    Carruthers, G.R. {\it et al.} 1980 {\it Astrophys. J.} 
{\bf 237}, 438. 

\bibitem{}    Carruthers, G. {\it et al.} 1993 {\it SPIE}
    {\bf 1764}, p. 21. 

\bibitem{}    Carruthers, G. {\it et al.} 1994 {\it SPIE} {\bf 2282}, p. 184.

\bibitem{}    Carruthers, G.R. \& Page, T. 1976 {\it Astrophys. J.} {\bf 205}, 397.

\bibitem{}    Carruthers, G.R. \& Page, T. 1983 {\it {\it Astrophys. J.} Suppl.} {\bf 53}, 623.

\bibitem{}    Carruthers, G.R. \& Page, T. 1984a {\it {\it Astrophys. J.} Suppl.} {\bf 54}, 271.

\bibitem{}    Carruthers, G.R. \& Page, T. 1984b {\it {\it Astrophys. J.} Suppl.} {\bf 55}, 101. 
  

\bibitem{} Carruthers, G.R. \& Seeley, T.D. 1996  SPIE
meeting no. 2831, Denver: August 7-8, 1996. 

\bibitem{}    Courtes, G. {\it et al.} 1984
    {\it Science} {\bf 225}, 179. 

\bibitem{}    Courtes, G. {\it et al.} 1993 {\it Astron. Astrophys.} {\bf 268}, 419.

\bibitem{} Courtes, G. {\it et al.} 1995 {\it Astron. Astrophys.} {\bf 297}, 338. 

\bibitem{}    Deharveng, J.-M. {\it et al.} 1979 {\it Space Science Instrumentation}
 {\bf 5},    21. 

\bibitem{} Deharveng, J.-M. {\it et al.} 1994, 
{\it Astron. Astrophys.} {\bf 289}, 715.

\bibitem{}    Donas, J. {\it et al.}
    1987, {\it Astron. Astrophys.} {\bf 180}, 12. 

\bibitem{}   Donas, J. {\it et al.} 1981 {\it Astron. 
Astrophys.} {\bf 97}, L7. 

\bibitem{} Donas, J. {\it et al.} 1995 {\it Astron. Astrophys.} {\bf 303}, 661. 

 
\bibitem{} Haikala, L.K. {\it et al.} {\it Astrophys. J. Lett.} {\bf 443}, L33.

\bibitem{}    Haisch, B. {\it et al.} 1976 {\it Astrophys. J.} {\bf 213}, L119.
 
\bibitem{} 	Heffernan, K.J. {\it et al.} 1996 {\it Johns Hopkins APL Technical
	Digest} {\bf 17}, 198. 

\bibitem{}    Henry, P. {\it et al.}
    1975a {\it Astrophys. J.} {\bf 195}, 107. 

\bibitem{}    Henry, P. {\it et al.}
    1975b {\it Rev. Sci. Instr.} {\bf 46}, 355.

 \bibitem{}   Henry, P. {\it et al.}
    1975c  {\it Astrophys. J. Lett.} {\bf 197}, L117.

\bibitem{}    Henry, R.C. 1982 {\it 10-th Texas Symposium, Ann
              NY Acad. Sci.,} p. 428.

\bibitem{}    Henry, R.C. 1991 {\it Ann. Rev.  Astron. Astrophys.} {\bf 29}, 89.

\bibitem{}    Hill, J.K. {\it et al.} 1993 {\it Astrophys. J.} {\bf 413}, 604. 

\bibitem{}    Hill, J.K. {\it et al.} 1994 {\it Astrophys. J.} {\bf 425}, 122. 

\bibitem{}    Hill, J.K. {\it et al.} 1995a {\it  Astrophys. J. Suppl.}  {\bf 98}, 595. 

\bibitem{}    Hill, J.K. {\it et al.} 1995b {\it Astrophys. J.} {\bf 446}, 622. 

\bibitem{}    Hill, J.K. {\it et al.} 1996  {\it Astrophys. J. Lett.} preprint.
 
\bibitem{}   Holberg, J.B.   1990 in "Observatories in Earth Orbit and Beyond", p. 49. 

\bibitem{}   Holberg, J.B.   1991 in "Extreme Ultraviolet Astronomy", p.8. 

\bibitem{}    Huguenin, D. \& Magnan, A. 1978  {\it ESA SP-135}, 403. 

\bibitem{}    Hurwitz, M. {\it et al.} 1991 {\it Astrophys. J.} {\bf 372}, 167. 

\bibitem{} Israel, D.J. 1993  {\it IEEE AES Systems Magazine (Feb. 1993),} p. 43.

 
\bibitem{}    Jelinsky, P. {\it et al.} 1995 {\it Astrophys. J.} 
{\bf 442}, 653.

 
\bibitem{} Kinney, A.L. {\it et al.} 1996 Astrophys. J. {\bf 467}, 38.

\bibitem{}    Kodaira, K. {\it et al.} 1990 {\it Astrophys. J.} {\bf 363}, 422. 

\bibitem{}    Laget, M. 1980 {\it Astron. Astrophys.} {\bf 81}, 37. 

\bibitem{}    Laget, M. {\it et al.} 1991a {\it Adv. Space
    Res.} {\bf 11}, 139. 

\bibitem{}    Laget, M. {\it et al.} 1991b {\it Astron.
    Astrophys.}   {\bf 259}, 510.

\bibitem{}    Lampton, M. {\it et al.}
    1976  {\it Astrophys. J. Lett.} 203, L71 

\bibitem{}    Lampton, M. {\it et al.} 1996 preprint. 

\bibitem{}    Lipovetsky, V.A. 1992 in "Astronomy from Wide-Field Imaging" (H.T.
    MacGillivray {\it et al.} eds.), Dordrecht: Reidel, p. 3.

 
\bibitem{}    Margon, B. {\it et al.}
    1976  {\it Astrophys. J. Lett.} {\bf 210}, L79.

\bibitem{}    Margon, B. {\it et al.}
    1978 {\it Astrophys. J.} {\bf 224}, 167.

 
\bibitem{}    Milliard, B. {\it et al.} 1983  {\it ESA SP-183}, p. 409.

\bibitem{}    Milliard, B. {\it et al.} 1991 {\it Adv. Space Res.} {\bf 11}, 135. 

\bibitem{}    Milliard, B. {\it et al.} 1992 {\it Astron. Astrophys.}
    {\bf 257}, 24.

\bibitem{} Murthy, J. {\it et al.} 1989 {\it Astrophys. J.}  {\bf 336}, 954.

\bibitem{}    Murthy, J. \& Henry, R.C. 1995 {\it Astrophys. J.} {\bf 448}, 848. 

 
\bibitem{}    O'Connell, R.W. 1987  {\it Astron. J.} {\bf 94}, 876 

 
\bibitem{}    O'Connell, R.W. 1991 {\it Adv. Space Res.} {\bf 11}, 71.  

 \bibitem{}    Onaka, T. {\it et al.} 1989 {\it Astrophys. J.} {\bf 342}, 235. 

\bibitem{}    Page, T. {\it et al.} 1982  {\it NRL Report} {\bf 8487}.  
 
\bibitem{}    Petit, H. {\it et al.} 1996 {\it Astron. Astrophys.} {\bf 309}, 446. 

 \bibitem{}   Pounds, K. \& Wells, A.A. 1991 {\it Adv. Space Res.} {\bf 11}, 125.

\bibitem{}    Pounds, K.A.  {\it et al.}  1993 {\it Mon. Not. R. astr. Soc.} {\bf 260}, 77.   

\bibitem{}    Priedhorsky, W.  {\it et al.}  1991 in "Extreme Ultraviolet Astronomy",  p. 464.

\bibitem{}    Pye, J.P. 1995 in "Astrophysics in the Extreme Ultraviolet"' proceedings
    of IAU Colloquium no. 152. 

\bibitem{}    Pye, J.P. {\it et al.} 1995 {\it Mon. Not. R. astr. Soc.}
    {\bf 274}, 1165. 

\bibitem{}    Reichen, M. {\it et al.} 1994
    {\it Astron. Astrophys. Suppl.} {\bf 106}, 523.

\bibitem{}    Sandel, B.R. {\it et al.} 1979, {\it Astrophys. J.} 
{\bf 227}, 808. 

\bibitem{}    Schmidt, E.G. \& Carruthers, G.R. 1993a {\it Astrophys. J.} {\bf 408}, 484.

\bibitem{}    Schmidt, E.G. \& Carruthers, G.R. 1993b {\it {\it Astrophys. J.} Suppl.} 
{\bf 89}, 259.

 \bibitem{}   Schmidt, E.G. \& Carruthers, G.R. 1995 {\it {\it Astrophys. J.} Suppl.} 
{\bf 96}, 605. 

\bibitem{}    Smith, A.M. \& Cornett, R.H. 1982 {\it Astrophys. J.} {\bf 261}, 1.

\bibitem{}    Smith, A.M. {\it et al.} 1987 {\it Astrophys. J.} {\bf 320}, 609. 
 

\bibitem{}    Smith, E.P. {\it et al.} 1996 {\it {\it Astrophys. J.} Suppl.} {\bf 104}, 287.  
 
\bibitem{}    Terebizh, V.Yu. 1986 {\it Contributions of Special Astrophys. Obs.} {\bf 50}, 79. 

\bibitem{}    Thompson, G.I. {\it et al.} 1978 "Catalog of Stellar Ultraviolet
    Fluxes", SRC. 

\bibitem{}    Tovmassian, H.M. {\it et al.} 1988 {\it Pis'ma v Astron. Zh.} {\bf 14}, 291 
                          (Sov. Astron. Lett. {\bf 14}, 123).

\bibitem{}    Tovmassian, H.M. {\it et al.} 1991a in "Astronomy
    from Wide-Field Imaging", p. 55. 

\bibitem{}    Tovmassian, H.T. {\it et al.}
    1991b in "Astronomy from Wide-Field Imaging",  p. 473. 

\bibitem{}    Tovmassian, H.M. {\it et al.} 1994 
{\it Astrophys. Space Sci.} {\bf 213}, 175. 

\bibitem{}    Tovmassian, H.M. {\it et al.}
    1993b {\it Astron. Astrophys. Suppl.} {\bf 100}, 501. 

\bibitem{}    Tovmassian, H.M. {\it et al.} 1993a
     {\it Astron. J.} {\bf 106}, 627. 

\bibitem{}    Tovmassian, H.M. {\it et al.} 1996a  {\it Astron. J.} {\bf 111}, 299. 

\bibitem{}    Tovmassian, H.M. {\it et al.} 1996b  {\it Astron. J.} 
{\bf 111}, 306. 

\bibitem{}    Van Duinen, R.J. {\it et al.} 1975 {\it Astron. Astrophys.} {\bf 39}, 159. 

\bibitem{}    Vuillemin, A. {\it et al.} 1991 {\it Adv. Space Res.} {\bf 11}, 143.
 

\bibitem{}    Waller, W.H. {\it et al.} 1995  {\it Astron. J.} {\bf 110}, 1255.

 
\end{thebibliography}
\end{document}